\documentclass{PoS}

\title{Charged Higgs Boson Physics\\ at Future Linear Colliders}

\ShortTitle{Charged Higgs Bosons at Linear Colliders}

\author{\speaker{Marco Battaglia}\thanks{current address CERN, Geneva, Switzerland.}\\
        SCIPP, University of California at Santa Cruz and CERN\\
        E-mail: \email{Marco.Battaglia@cern.ch}}

\abstract{The search and study of heavy Higgs bosons is an important part of the 
anticipated physics program of a high energy $e^+e^-$ linear collider. This paper
reviews the expected sensitivity of $e^+e^-$ collisions from 0.5 up to 3~TeV in the 
study of the physics of charged Higgs bosons in supersymmetric scenarios.}

\FullConference{Third International Workshop on Prospects for Charged Higgs Discovery 
at Colliders - CHARGED2010,\\
September 27-30, 2010\\
Uppsala Sweden}

\begin{document}

\section{Introduction}

The study of an extended Higgs sector appears particularly adapted to the capabilities of 
a lepton collider of sufficient energy and luminosity. $e^+e^-$ and $\mu^+ \mu^-$ collisions 
provide us with pair production of $H^+H^-$ and $H^0 A^0$ bosons. If photon collisions are 
available in a $\gamma \gamma$ collider setup, scans of the $H^0$ and $A^0$ peaks can also 
be performed~\cite{Illana}. In this paper we focus on the study of charged Higgs bosons at 
an $e^+e^-$ linear collider in supersymmetric scenarios. 
Since the heavy Higgs states generally decouple from the other Supersymmetric particles, 
most of, if not all, the results derived for the Minimal Supersymmetric extension of the 
Standard Model (MSSM) can be also applied to a non-Supersymmetric extended Higgs sectors, 
such as that of the two-Higgs doublet model (2HDM) or the Higgs triplet model~\cite{huitu}. 
In the decoupling limit, with $M_A >> M_Z$ and the 
masses of the heavy Higgs particles much larger than those of the other states, the fundamental 
quantities to be determined in the charged Higgs study are the masses and the widths of these Higgs 
states and their decay branching fractions. The mass of one of the heavy Higgs bosons, generally 
taken to be the CP-odd $A^0$ state, is a fundamental theory parameter. The $A^0$, $H^0$ and $H^{\pm}$ 
bosons are almost degenerate in mass in the decoupling limit and this needs to be verified 
experimentally. 
The heavy Higgs decay branching ratios are sensitive to another fundamental theory parameter, 
the ratio of the vacuum expectation values of the two Higgs doublets $\tan \beta = v_2/v_1$.
The total width of the heavy Higgs states is also important for determining $\tan \beta$.
Finally, in SUSY models the contribution of sparticle loops may induce CP-violating asymmetries 
in the decay of charged Higgs bosons, which can be searched for at a linear collider.
From an experimental point of view, the large jet multiplicity and event complexity of 
processes such as $e^+e^- \to H^+H^- \to t \bar b \bar t b$ and $W^+ h^0 W^- h^0$ make them
excellent benchmarks for the detector granularity and its ability to perform accurate kinematic 
reconstruction and jet flavour tagging in an high particle density environment. 

\section{Charged Higgs Production and Simulation Studies}

\begin{figure}[h!]
\begin{center}
\includegraphics[width=7.0cm]{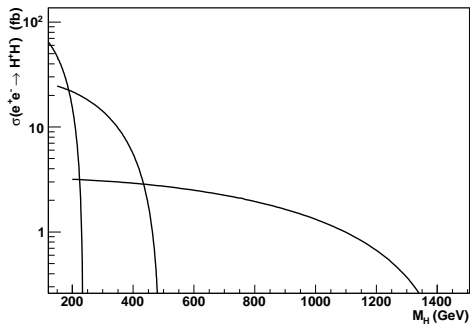}
\end{center}
\caption{Higgs pair production: cross section for $e^+e^- \to H^+H^-$ at $\tan \beta$ = 20 as a 
function of $M_{H^{\pm}}$ for $\sqrt{s}$ = 0.5, 1.0 ad 3.0~TeV.}
\label{fig:xsec}
\end{figure}

Charged Higgs production in $e^+e^-$ collisions proceeds mostly through s-channel pair production, 
$e^+e^- \to H^+H^-$. The production cross section at a given $\sqrt{s}$ energy depends on the 
charged Higgs mass and $\tan \beta$. Typical values range between $\sim$20~fb for $M_{H^{\pm}}$ = 
250~GeV at $\sqrt{s}$ = 0.8~TeV and 0.5~fb for $M_{H^{\pm}}$ = 1140~GeV at $\sqrt{s}$ = 3~TeV,  
due to the P-wave suppression near threshold (see Figure~\ref{fig:xsec}). 
The single boson production $e^+e^- \to H^- \tau^+ \nu_{\tau}$ + c.c.\ \cite{kanemura,moretti} 
and $e^+e^- \to H^- t \bar b$ + c.c.\ \cite{kniehl} processes, which give access to charged Higgs 
production beyond the kinematic limit for pair production, have also been considered.  Their cross 
sections scale as $\tan^2 \beta$. However, at $\cal{O}$(0.01~fb) they are too small to be exploitable 
in the interesting region $M_{H^{\pm}} > \sqrt{s}/2$. Production of charged Higgs bosons in $t$ decays,
$t \to H^+ b$ has also been studied at a linear collider, using also the fact that the polarisation of 
the tau leptons from the subsequent $H^{\pm} \to \tau^{\pm} \nu_{\tau}$ decay is opposite to that of 
those originating from $W^{\pm} \to \tau^{\pm} \nu_{\tau}$ for isolating the signal~\cite{boos}. However, 
the Tevatron~\cite{cdf,d0}, and soon the LHC, data are closing the region of interest for this 
process.   


In the mass range of interest for a linear collider, the main decay processes are  $H^+ \to t \bar b$, 
$H^+ \to \tau^+ \bar \nu_{\tau}$ and $H^+ \to W^+ h^0$. The $t \bar b \bar t b$ final state is 
dominant and it has been considered for most of the searches and mass measurement studies. 
Simulation studies of charged Higgs bosons have been performed at 0.5~TeV for $M_{H^{\pm}}$= 140 
and 180~GeV~\cite{sopczak}, at 0.8~TeV for $M_{H^{\pm}}$= 200 and 300~GeV~\cite{kiiskinen, tesla}
and at 3.0~TeV for $M_{H^{\pm}}$= 700~\cite{Coniavitis:2007me,Coniavitis:2008zz}, 900~\cite{clicphys} 
and 1140~GeV. The SM $t \bar b \bar t b$ irreducible background is small, with a production cross 
section of 3.2~fb at 0.8~TeV and 1.2~fb at 3~TeV and a flat distribution in the $tb$ invariant mass. 
The event reconstruction starts with the identification of two top quarks in events with large jet 
multiplicity and no significant missing energy. At energies below 1 TeV the boost of the top is 
small and the decay $t \to W^+ b$ can be explicitly reconstructed. At multi-TeV energies and 
$M_{H^{\pm}}$ of 0.7 - 1.2~TeV, the top quark boost is such that it can be reconstructed as 
a single jet. Jet flavour tagging is essential since the four $b$ jets are a distinctive 
signature of the $WbbWbb$ final state. Once the four-parton final state is reconstructed, the $t$ 
and $b$ pair which minimises the mass difference between the two $tb$ systems is selected. Given 
the large jet multiplicity, the four $b$ tags and the need to reject leptonic $W$ decays for an 
accurate $t$ energy measurement, the reconstruction efficiency is of the order of just a few percent. 
\begin{figure}[hb!]
\begin{center}
\begin{tabular}{cc}
\includegraphics[width=5.5cm]{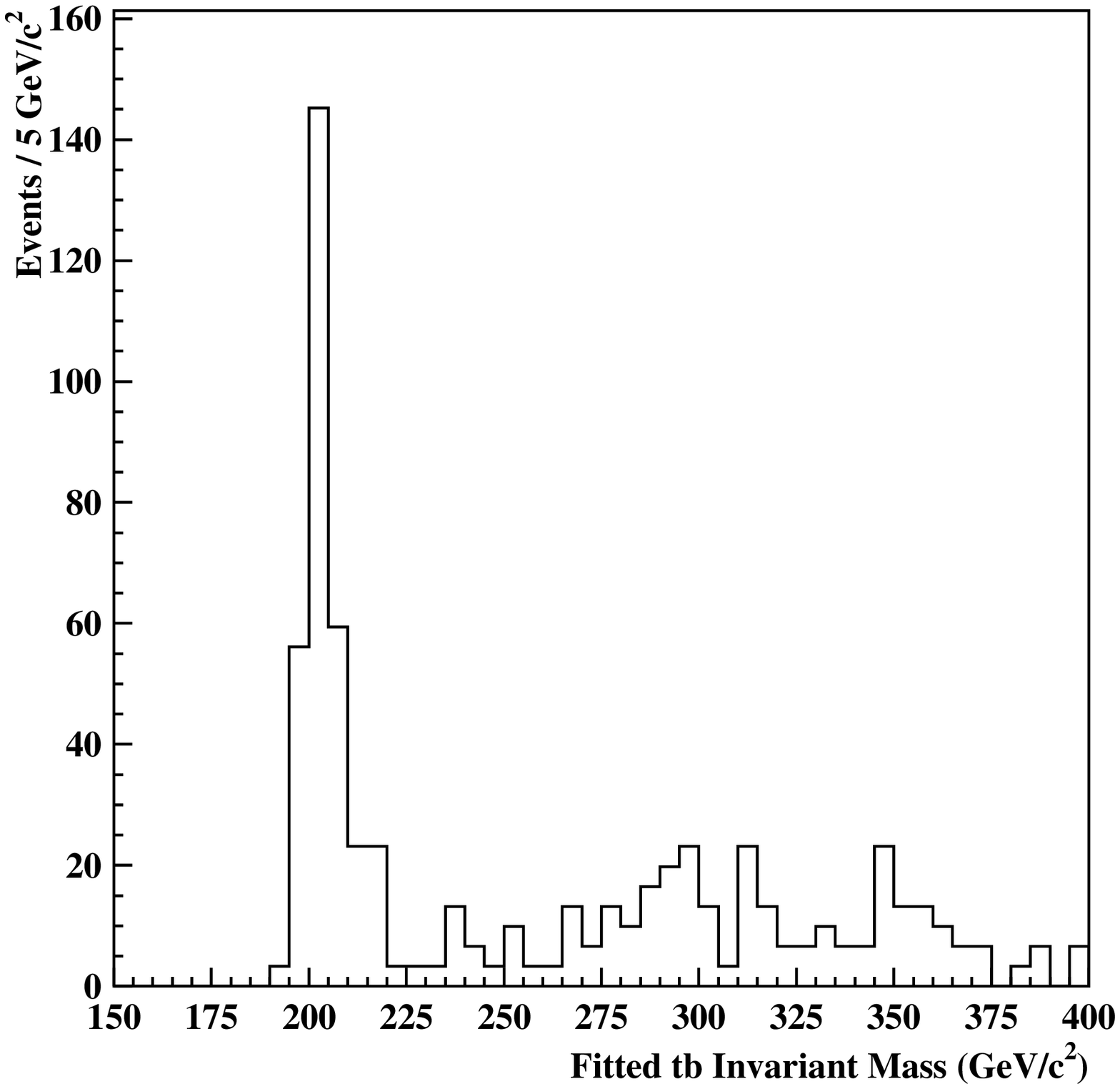} &
\includegraphics[width=5.5cm]{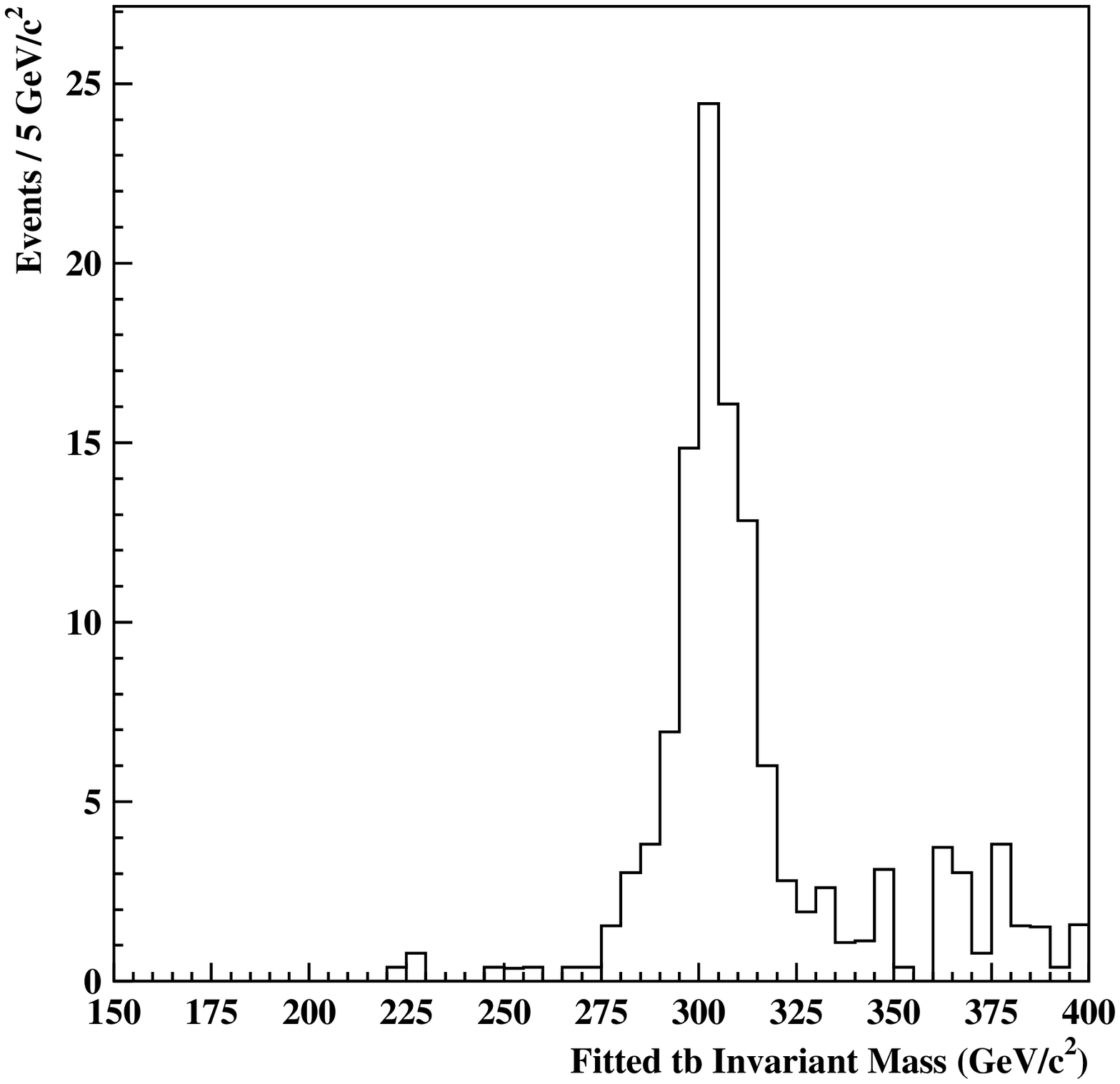} \\
\end{tabular}
\end{center}
\caption{Mass reconstruction: $tb$ invariant mass after kinematic fitting for signal events with 
$M_{H^{\pm}}$ = 200~GeV (left) and 300~GeV (right) at $\sqrt{s}$ = 0.8~TeV (from~\cite{kiiskinen}).}
\label{fig:mass}
\end{figure}

The $tb$ invariant mass resolution is 16~GeV for $M_{H^{\pm}}$ = 180~GeV at $\sqrt{s}$ = 
0.5~TeV~\cite{sopczak}, 17~GeV for $M_{H^{\pm}}$ = 300~GeV at $\sqrt{s}$ = 0.8~TeV~\cite{kiiskinen}, 
55~GeV for $M_{H^{\pm}}$ = 700~GeV~\cite{Coniavitis:2007me,Coniavitis:2008zz} and 72~GeV for 
$M_{H^{\pm}}$ = 1140~GeV, both at $\sqrt{s}$ = 3.0~TeV.
The mass resolution is due to the finite jet energy resolution, to missing energy from escaping neutrinos
and to confusion in the association of individual particles to jets, due to the large multiplicities.
This mass resolution can be improved by applying a kinematic fit to the reconstructed hadronic final 
states. Kinematic fits applied in simulation studies  impose energy and transverse momentum 
$p_x$ = $p_y$ = 0, $E \pm |p_z| = \sqrt{s}$, accounting for beamstrahlung photons 
radiated along the beam axis and an equal boson mass condition $M_{j1j2} = M_{j3j4}$.
This improves the mass resolution by a factor of $\sim$2 to 8.5~GeV for $M_{H^{\pm}}$ = 300~GeV 
at $\sqrt{s}$ = 0.8~TeV (see Figure~\ref{fig:mass})~\cite{kiiskinen}, 25~GeV for $M_{H^{\pm}}$ = 
700~GeV at $\sqrt{s}$ = 3.0~TeV~\cite{Coniavitis:2007me,Coniavitis:2008zz} and 45~GeV for 
$M_{H^{\pm}}$ = 1140~GeV. However, these resolutions are larger than the expected $H^{\pm}$
natural width at these masses.

\section{Discovery Reach and Present Predictions}

\begin{figure}[b!]
\begin{center}
\includegraphics[width=8.5cm,clip=]{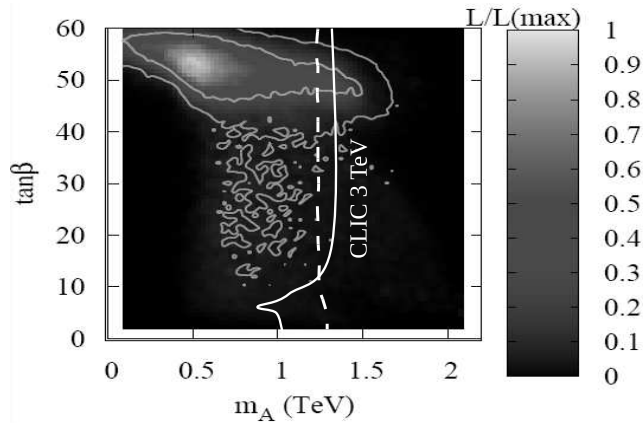}
\end{center}
\caption{Likelihood marginalised to the $M_A$ - $\tan \beta$ plane with the contour for the
 anticipated sensitivity to $H^+H^-$ (dashed line) and $H^0A^0$ (continuous line) pair production 
at a 3~TeV linear collider (plot modified from~\cite{Allanach:2005kz}, sensitivity contours 
from~\cite{Coniavitis:2007me})}
\label{fig:matanb}
\end{figure}

Since the SM background is small and flat at large values of the $tb$ invariant mass, 
a significant signal of charged Higgs pair production can be established at an $e^+e^-$ linear 
collider, even with small cross sections and the sensitivity extends almost to the kinematical 
threshold for pair production by combining different decay final states, irrespective of the value 
of $\tan \beta$. At $\sqrt{s}$ = 3~TeV, the $e^+e^- \to H^+H^-$ process is observable for charged 
Higgs boson masses up to 1.25~TeV (see Figure~\ref{fig:matanb}) and exceeds the sensitivity of 
the process $e^+e^- \to H^0A^0$ which drops below 1~TeV at $\tan \beta \sim$ 5~\cite{Coniavitis:2007me}.
This sensitivity can be compared to the expectations for the heavy Higgs mass values 
of new physics models within the constraints offered by previous searches, low energy 
data and, possibly, astrophysical data, such as relic dark matter density. There have 
been extensive efforts to determine the regions of new physics model parameter space most 
likely to be realised in nature given these data. Most of these efforts concern some 
specific implementation of the supersymmetric extension of the Standard Model. The 
constraints most relevant here are those from the light Higgs boson mass limits, rare 
$B$ decay rates and dark matter relic density. Given the large number of parameters of 
the MSSM, these studies are best performed in constrained versions, where the gaugino 
and sfermion masses are assumed to be unified at the GUT scale. This is generally known 
as the constrained MSSM (cMSSM). The physical masses are calculated using the 
renormalisation group equations (RGEs) from a set of five fundamental parameters: the universal 
scalar, $m_0$, and gaugino, $m_{1/2}$, masses, $\tan \beta$, the universal trilinear coupling, 
$A_0$ and the sign of the coupling $\mu$. Different statistical techniques have been employed. 
A sampling of the full five-parameter space was performed in the study of ref.~\cite{Allanach:2005kz} 
using a Markov chain Monte Carlo method~\cite{baltz} and taking into account limits from 
particle searches, $\Omega_{\chi} h^2$ and low-energy constraints. Of particular interest 
here is the two-dimensional map of the $M_A$ - $\tan \beta$ plane obtained, which is shown 
in Figure~\ref{fig:matanb} where the sensitivity of a 3~TeV linear collider to $H^{\pm}$ and 
$H^0A^0$ pair production is superimposed.
Variants of the cMSSM either add further constraints or relax them. In mSUGRA there are fixed 
relations between the bi-linear and trilinear couplings parameters and the scalar mass and the 
additional condition that the gravitino mass is equal to the scalar mass, $m_{3/2} = m_0$, 
thus reducing the number of free parameters to just three~\cite{Cremmer:1978iv}. 
In the non-universal Higgs mass model (NUHM) the supersymmetric contributions to the Higgs 
masses are allowed a different value~\cite{Baer:2005bu}. In this model, predictions for the 
$M_A$-$\tan \beta$ parameters have been obtained in the study of ref.~\cite{Ellis:2007}.
A recent frequentist analysis, based on LEP-2 limits, electro-weak and flavour data, $g-2$ and 
$\Omega_{\chi} h^2$ results, considers all these models and determines likelihood functions for 
SUSY observables. The 95~\% C.L. interval for the charged Higgs boson mass is found to be in the 
range 350 $< M_{H^{\pm}} <$ 750~GeV for the cMSSM and in the range 100 $< M_{H^{\pm}} <$ 600~GeV 
for the non-universal Higgs mass extension~\cite{ellis-cmssm}. A similar study in the mSUGRA 
model finds the interval 600 $< M_{H^{\pm}} <$ 1200~GeV, again at 95~\% C.L.~\cite{ellis-msugra}. 
These results show that present data place only a very loose constrain on the masses of the heavy 
Higgs particles and highlight the interest in pursuing detailed studies for charged Higgs bosons 
at the LHC and at future lepton colliders over a broad mass range up to, and possibly beyond, 1~TeV. 

\subsection{Mass Determination}

\begin{figure}
\begin{center}
\begin{tabular}{cc}
\includegraphics[height=6.0cm,clip=]{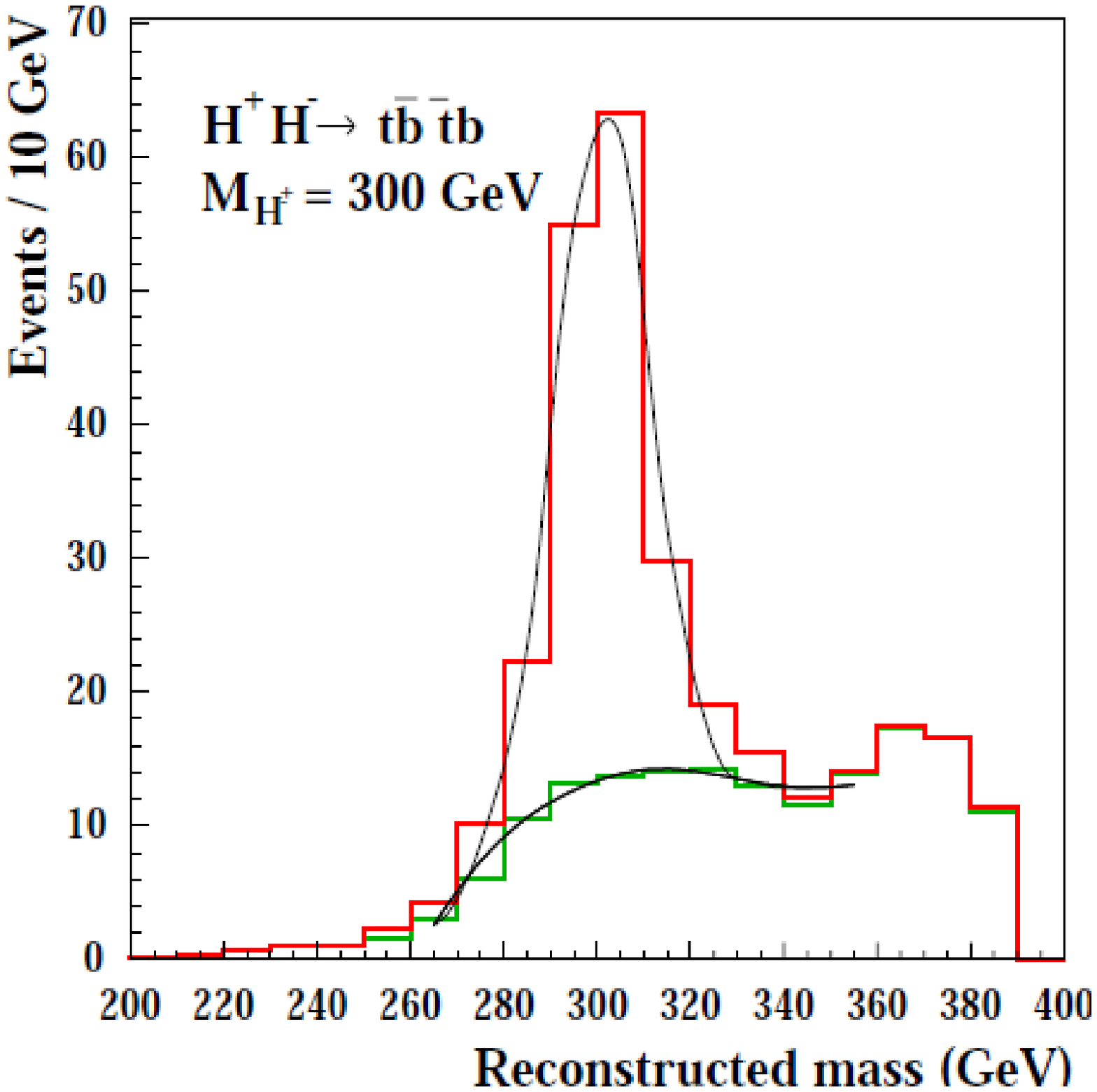} &
\includegraphics[height=6.0cm,clip=]{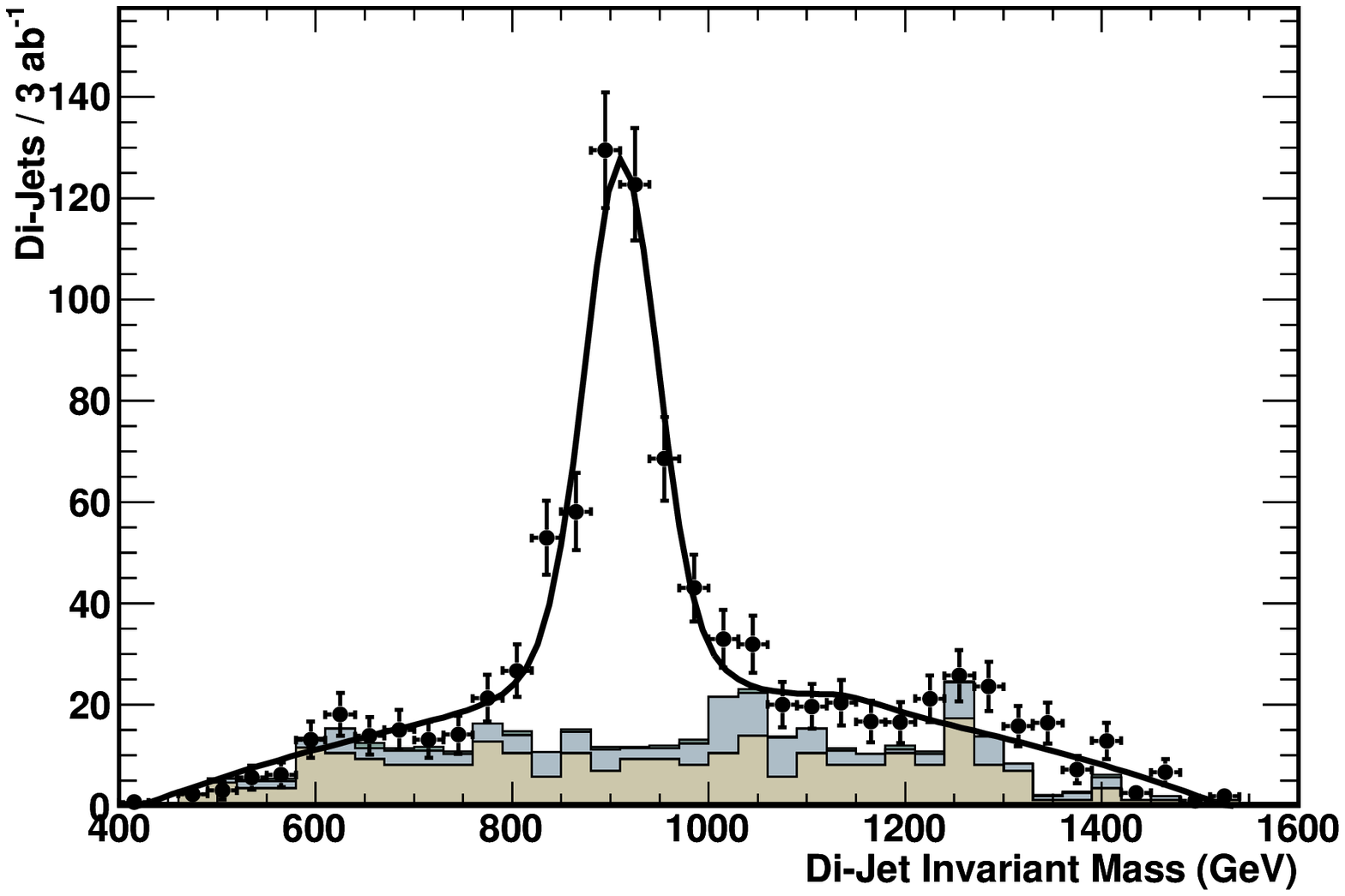} \\
\end{tabular}
\end{center}
\caption{Mass reconstruction: $tb$ invariant mass for signal and background events reconstructed 
at $\sqrt{s}$ = 0.8~TeV (left) (from~\cite{tesla}) and 3.0~TeV (right).}
\label{fig:mass2}
\end{figure}

The mass $M_{H^{\pm}}$ is determined by a fit to the two-parton invariant mass distribution after 
kinematic fitting. 
The signal peak can be modelled by a Breit-Wigner function convoluted with a Gaussian resolution 
term. In general, the experimental resolution is larger than the $H^{\pm}$ natural width. 
Alternatively, a template method, as that adopted in the study of ref.~\cite{Coniavitis:2007me}, 
can be used to build a $\chi^2$ for the mass fit. The SM background is essentially flat below this 
peak in the mass region of interest. The residual $H^0A^0 \to b \bar b \bar b b$ contribution gives 
a peaking background in the signal region, since the masses of the charged and heavy neutral Higgs 
bosons are expected to be almost degenerate. However, the $e^+e^- \to H^+H^-$ cross section is larger 
than that for $e^+e^- \to H^0A^0$ and the $t$ tagging reduces this background. 
The invariant mass distributions for signal and background at $\sqrt{s}$ of 0.8~TeV and 3.0~TeV are 
shown in Figure~\ref{fig:mass2}.

\begin{table}
\caption{Summary of the results on the reconstruction efficiency and the determination of $M_{H^{\pm}}$ 
at various centre-of-mass energies.}
\begin{center}
\begin{tabular}{|c|c|c|c|c|c|c|}
\hline
 $M_{H^{\pm}}$ & $\sqrt{s}$ & $\int \cal{L}$ & Final & Selection  & $\delta M / M$ & Ref.\ \\
   (GeV)       & (TeV)      & (ab$^{-1}$)    & State & Efficiency &                &       \\
\hline
~145            & 0.5        & 0.5            & $cs$ & 0.15       & 0.006         & \cite{sopczak}   \\
~200            & 0.8        & 0.5            & $tb$ & 0.02       & 0.002         & \cite{kiiskinen,tesla} \\
~300            & 0.8        & 0.5            & $tb$ & 0.04       & 0.004         & \cite{kiiskinen,tesla} \\
~702            & 3.0        & 3.0            & $tb$ & 0.02       & 0.007         & \cite{Coniavitis:2007me}\\
~880            & 3.0        & 3.0            & $tb$ & 0.05       & 0.008         & \cite{clicphys}\\
~906            & 3.0        & 3.0            & $tb$ & 0.07       & 0.006         & Preliminary     \\
1136            & 3.0        & 3.0            & $tb$ & 0.05       & 0.007         & Preliminary     \\
\hline
\end{tabular}
\end{center}
\label{tab:mass}
\end{table}

\subsection{Decay Studies}

Modes other than the dominant $tb$ decay have also been studied. In particular, the 
study of~\cite{Coniavitis:2007me} has considered the leptonic decay $H^+ \to \tau^+ \nu_{\tau}$ 
in the mixed channel $t b \tau \nu$. Signal events can be selected using the standard reconstruction 
of the hadronic decay and the transverse mass of $H^+ \to \tau^+ \nu_{\tau}$.
The reconstruction efficiency, around 0.045, is comparable to that of the fully hadronic mode.
The determination of $\tan \beta$ will be essential to constrain phenomenology and to relate 
the extended Higgs sector to cosmology through dark matter. The precision of a linear 
collider is expected to provide essential inputs~\cite{Baltz:2006fm}.
It is possible to constrain $\tan \beta$  by determining the $H^0$, $A^0$, $H^+$ widths and 
decay branching ratios $H^0$, $A^0 \to b \bar b$, $\tau^+ \tau^-$ , $H^+ \to t \bar b$, $\tau \nu$. 
In particular, the determination of the charged Higgs bosons decay yields to $tb$ and $cs$ 
hadronic and to $\tau \nu_{\tau}$ leptonic final states offers a good opportunity to determine
$\tan \beta$ as long as its value is not too large~\cite{Gunion:2002ip,Coniavitis:2007me}. 
\begin{figure}
\begin{center}
\includegraphics[width=6.5cm,angle=90]{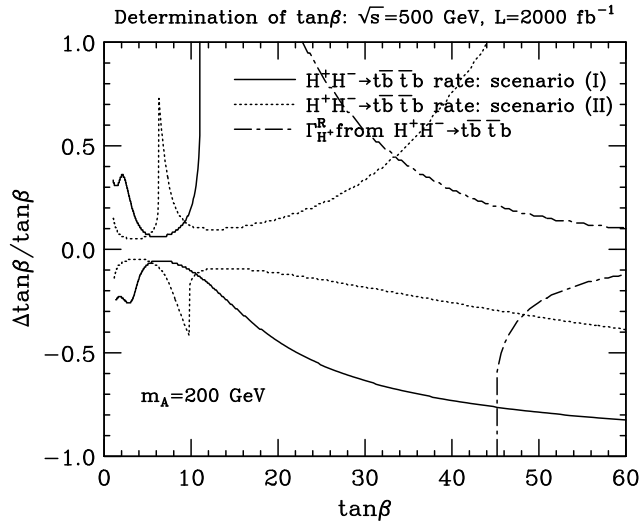}
\vspace*{-0.25cm}
\end{center}
\caption{$\tan \beta$ determination with $H^{\pm}$: accuracy as a function of $\tan \beta$ 
for $M_{H^{\pm}}$ = 200~GeV (from~\cite{Gunion:2002ip})}
\label{fig:tanb}
\end{figure}
At very large values of $\tan \beta$ complementary sensitivity is provided by the 
determination of the boson width as shown in Figure~\ref{fig:tanb}.

In presence of CP violation, the scalar $h^0$ and $H^0$ bosons and the pseudo-scalar $A^0$ 
are mixed, In this case the $A^0$ is no longer a mass eigenstate and the $H^{\pm}$ mass
should be used instead of $M_A$ for parametrising the MSSM~\cite{ellis-cpv}. 
CP violation in the new physics sector can manifest itself in decay rate asymmetries such as 
\begin{equation}
\delta^{CP}_{f \bar f^{\prime}} = \frac{\Gamma(H^+ \to f \bar f^{\prime}) - 
 \Gamma(H^- \to \bar f f^{\prime})}
{\Gamma(H^+ \to f \bar f^{\prime}) + \Gamma(H^- \to \bar f f^{\prime})}
\end{equation}
In particular, in MSSM  CP asymmetries are expected to be mostly due to squark loops in 
$H^+ \to t \bar b$ decays~\cite{Christova:2002sw,Christova:2002ke,Christova:2006fb}. The size of 
these asymmetries scale inversely with $\tan \beta$ from 0.20 to 0.02 for 5 $< \tan \beta <$ 30. 
$\delta^{CP}$ can be determined using quark-anti quark tagging in hadronic top decays adopting 
a vertex charge algorithm for $b$ and $c$ jets and lepton charge in semileptonic $t$ decays. 
The analysis is quite challenging due to the limited quark charge separation power and the small 
statistics. With typical cross sections of ${\cal{O}}$(1~fb) and few ab$^{-1}$ of integrated 
luminosity, a sensitivity to deviations of $\delta^{CP}$ from zero at the level of $\sim$0.10 
should be feasible.

\section{Conclusions}

A lepton collider of sufficiently high energy and luminosity can provide good accuracy
in the determination of the mass (better than 1\%), production cross section and decay 
branching fractions of charged Higgs bosons, through pair production virtually up to the 
kinematic limit.
The accuracy afforded by an $e^+e^-$ linear collider will be essential for the interpretation 
of the nature of a non-minimal Higgs sector and understand its role in relation to relic dark 
matter. The study of the heavy Higgs sector is one of the important drivers towards high energy 
and high luminosity performance for a linear collider. At present the input from the LHC data 
is essential to define the machine and detector requirements. The ongoing activity in simulation 
studies for the ILC at 1~TeV and CLIC at 3 TeV will help clarifying requirements and physics 
potential based on realistic simulation and reconstruction including accelerator effects.

\section*{Acknowledgements}

It is a pleasure to thank Tord Ekelof for the kind invitation to talk at this workshop
and Arnaud Ferrari for useful discussion.

\end{document}